# Hybrid Quantum-HPC Solutions for Max-Cut: Bridging Classical and Quantum Algorithms

1st Ishan Patwardhan
*Department of Computational Sciences*
*COEP Technological University*
Pune, India
patwardhanip20.comp@coeptech.ac.in

2nd Akhil Akkapelli
*Robert Bosch Centre for Cyber-Physical Systems*
*Indian Institute of Science*
Bengaluru, India
aakhil@iisc.ac.in

*Abstract*—This research explores the integration of the Quantum Approximate Optimization Algorithm (QAOA) into hybrid quantum-HPC systems for solving the Max-Cut problem, comparing its performance with classical algorithms like brute-force search and greedy heuristics. We develop a theoretical model to analyze the time complexity, scalability, and communication overheads in hybrid systems. Using simulations, we evaluate QAOA's performance on small-scale Max-Cut instances, benchmarking its runtime, solution accuracy, and resource utilization. The study also investigates the scalability of QAOA with increasing problem size, offering insights into its potential advantages over classical methods for large-scale combinatorial optimization problems, with implications for future quantum computing applications in HPC environments.

## I. INTRODUCTION

Combinatorial optimization problems, such as the Max-Cut problem, play a crucial role in various fields, including network design, statistical physics, and machine learning [1]. These problems are computationally challenging because their solution spaces grow exponentially with the size of the input, making them intractable for classical algorithms, especially as the problem size increases [2]. One such classical algorithm, the brute-force approach, explores every possible solution, resulting in exponential time complexity. Although heuristic methods, such as greedy algorithms, offer faster approximations, they often fail to provide optimal solutions for large-scale problems [3].

The Quantum Approximate Optimization Algorithm (QAOA) has emerged as a promising candidate for solving combinatorial optimization problems more efficiently. QAOA leverages quantum mechanics' unique properties—such as superposition and entanglement—to explore solution spaces more effectively than classical algorithms [4]. By applying quantum circuits that alternate between the cost Hamiltonian and a mixing Hamiltonian, QAOA optimizes variational parameters to approximate solutions to problems like Max-Cut [5]. Notably, QAOA has shown potential for offering approximate solutions with lower computational complexity, especially as quantum hardware continues to advance [6]. In particular, QAOA has shown potential for providing approximate solutions to problems like Max-Cut with lower computational complexity in theory, making it a key area of interest in the development of quantum algorithms.

Despite advances in quantum computing hardware, the integration of quantum algorithms such as QAOA into classical hybrid quantum-HPC systems is still in its infancy. This integration introduces a range of challenges, including managing communication overhead, optimizing circuit depth, and ensuring efficient task scheduling between quantum and classical components [7]. Additionally, while there has been substantial progress in optimizing QAOA circuits for specific quantum architectures, there is a lack of comprehensive models that explain how QAOA can be integrated into hybrid quantum-HPC workflows and how its performance compares to that of classical algorithms. Moreover, existing models for QAOA are often hardware-dependent, and there is a need for more robust frameworks that integrate QAOA into hybrid quantum-HPC workflows to address large-scale combinatorial problems [8].

Recent studies have demonstrated the potential of recursive QAOA (RQAOA), which enhances QAOA by recursively solving subproblems, achieving superior performance on certain graph structures [9]. Additionally, the integration of classical approximation algorithms such as Goemans-Williamson with quantum approaches highlights the ongoing efforts to improve QAOA's performance [10].

This research aims to address these gaps by developing a theoretical model for the implementation of QAOA in hybrid quantum-HPC systems, focusing on the Max-Cut problem as a case study. Specifically, the research will compare the performance of QAOA with classical algorithms, such as brute-force search and greedy heuristics, to assess the advantages of using quantum computing in solving large-scale combinatorial optimization problems. Through both theoretical analysis and quantum circuit simulations, this study seeks to evaluate the scalability, complexity, and resource requirements of QAOA, while also investigating its potential to outperform classical methods within hybrid quantum-HPC systems.

## II. LITERATURE REVIEW

The Max-Cut problem is a prominent NP-hard combinatorial optimization problem that has attracted considerable attention due to its applications in fields such as physics, networking, and circuit design. Classical algorithms for solving Max-Cut include exact methods like brute-force search, which evaluates all possible partitions of the vertices. While brute-force guarantees an optimal solution, its exponential time complexity $O(2^n)$ limits its practical applicability to small graphs [11]. Greedy algorithms, though significantly faster with time complexity $O(E)$, where $E$ is the number of edges, often provide suboptimal solutions and struggle with larger, denser graphs. As the scale of these optimization problems grows, classical algorithms face a significant computational bottleneck.

More advanced classical approaches, such as the Goemans-Williamson algorithm, have been developed to approximate solutions to the Max-Cut problem. This semi-definite programming (SDP)-based algorithm provides a guaranteed approximation ratio of 0.878 and operates in polynomial time, offering a more scalable alternative to brute-force methods. However, SDP-based algorithms, while efficient for medium-sized problems, can also encounter scalability issues as problem sizes increase, leading to high computational overhead [12].

These limitations have motivated the exploration of quantum computing as a potential solution to address the computational complexity of Max-Cut and other combinatorial optimization problems. Quantum computing has emerged as a promising paradigm for solving optimization problems that are intractable for classical algorithms. The Quantum Approximate Optimization Algorithm (QAOA), introduced by Farhi, is designed specifically for combinatorial optimization and is suitable for current Noisy Intermediate-Scale Quantum (NISQ) devices [13]. QAOA operates by alternating between two parameterized quantum circuits: a cost Hamiltonian that encodes the optimization problem and a mixer Hamiltonian that explores the solution space by transitioning between different quantum states. The algorithm optimizes the parameters of these circuits to approximate a solution. Early experiments with QAOA have shown that it can outperform classical algorithms for specific instances of optimization problems. However, its performance is heavily dependent on the number of qubits, the depth of the quantum circuit, and the noise levels of the quantum hardware [14].

*A. Quantum vs Classical Approaches: A Deeper Comparison*

Although QAOA has proven effective, it is still essential to benchmark its performance against classical algorithms like brute-force and greedy heuristics in hybrid quantum-HPC environments. Studies have shown that hybrid systems can optimize the computational load by assigning classical and quantum tasks based on their respective strengths [15]. While classical algorithms are known to scale poorly with larger problem sizes, quantum algorithms, especially QAOA, have the potential to address these limitations. In particular, it has been demonstrated that QAOA can solve instances of Max-Cut more efficiently than classical methods when a sufficient number of qubits are available, though the exact quantum speedup is highly contingent on the depth of the quantum circuit [16].

*B. Optimization Techniques for QAOA Circuits*

Circuit optimization is essential for improving the practical implementation of QAOA. Given the limited coherence times and high noise levels in NISQ devices, reducing the depth and gate count of quantum circuits is crucial. The Maaps method, proposed by Zhu, optimizes QAOA circuits by reordering commuting gates, reducing the circuit depth while preserving the structure of the algorithm [17]. This method has been successfully applied to quantum architectures like Google Sycamore and IBM's heavy-hex grid, showing significant improvements in circuit fidelity and execution time. These circuit optimizations are critical for scaling QAOA to solve larger problem instances on near-term quantum hardware.

*C. Hybrid Quantum-Classical Systems and Task Allocation*

Integrating QAOA into hybrid quantum-HPC systems offers a promising approach to solving large-scale optimization problems. In a hybrid system, quantum processors can be used to solve specific subproblems or accelerate optimization tasks, while classical HPC systems handle pre- and post-processing, as well as other computationally intensive tasks. This integration enables the efficient division of labor between quantum and classical resources. Perdomo-Ortiz has demonstrated the potential of hybrid quantum-classical systems in optimizing certain combinatorial problems, showing that quantum devices can outperform purely classical methods in specific cases [18]. However, the successful implementation of such systems requires addressing several challenges, including efficient task scheduling, data transfer between classical and quantum components, and minimizing communication overhead.

*D. Challenges in Hybrid Quantum-HPC Environments*

Guerreschi studied the integration of variational quantum algorithms, including QAOA, within hybrid quantum-HPC environments [19]. His work highlighted several key challenges, such as the latency introduced by frequent communication between the classical and quantum components, and the need for efficient scheduling to avoid bottlenecks in hybrid workflows. Furthermore, his research emphasized the importance of optimizing circuit depth and task division to fully exploit the capabilities of hybrid systems. In a related study, Matsuura analyzed the impact of quantum-classical communication bandwidth on the overall performance of hybrid quantum systems, concluding that minimizing this overhead is essential for achieving scalable performance improvements [20].

While these studies have made significant progress, there is still a gap in research regarding the direct comparison of QAOA's performance with classical algorithms, such as brute-force and greedy heuristics, within a hybrid quantum-HPC framework. Current literature has mostly focused on either the optimization of QAOA circuits or the general integration of quantum computing into classical systems, without providing a detailed comparative analysis of how QAOA performs relative to classical methods in solving the Max-Cut problem within hybrid environments.

This research aims to fill this gap by developing a theoretical model that evaluates the performance of QAOA within a hybrid quantum-HPC system, specifically targeting the Max-Cut problem. We will compare the performance of QAOA against classical algorithms, including brute-force search and greedy heuristics, focusing on metrics such as execution time, solution accuracy, scalability, and resource utilization. By simulating QAOA circuits and modeling hybrid workflows, this study seeks to provide insights into the conditions under which QAOA outperforms classical approaches and to offer guidance on the practical implementation of QAOA in real-world hybrid systems.

## III. PROPOSED METHODOLOGY

This section outlines the methodology for modeling and implementing the *Quantum Approximate Optimization Algorithm (QAOA)* for solving the *Max-Cut problem* in the context of a hybrid quantum-classical computing environment. The methodology involves multiple components: problem formulation, the design and implementation of quantum circuits for QAOA, the formulation of classical algorithms for comparison, and the integration of these methods within hybrid quantum-high performance computing (HPC) systems.

### A. Problem Formulation: The Max-Cut Problem

The *Max-Cut problem* is a classical optimization problem in graph theory. Given a graph $G = (V, E)$, where $V$ represents the set of vertices and $E$ represents the set of edges, the goal is to partition the vertices into two disjoint sets $S_1$ and $S_2$ such that the number of edges between the two sets is maximized. This number of edges is referred to as the "cut" of the graph [21] as illustrated in Fig 1.

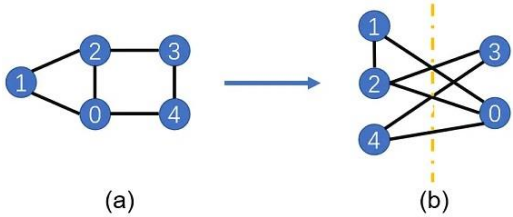

Fig. 1. Illustration of max cut problem

Mathematically, the Max-Cut problem can be represented by the following objective function:

$$C(S) = \sum_{\{u,v\} \in E} \frac{1}{2}(1 - z_u z_v) \quad (1)$$

where $z_u \in \{-1, 1\}$ represents the assignment of vertex $u$ to one of the two sets, and $z_v \in \{-1, 1\}$ represents the assignment of vertex $v$. If $z_u \neq z_v$, the edge between vertices $u$ and $v$ contributes to the cut. The challenge of solving Max-Cut stems from its combinatorial nature: the number of possible partitions grows exponentially with the size of the graph, making exact solutions computationally infeasible for large graphs [22].

### B. The Quantum Approximate Optimization Algorithm (QAOA)

*1) Overview of QAOA:* The *Quantum Approximate Optimization Algorithm (QAOA)* is a quantum algorithm designed to approximately solve combinatorial optimization problems, including Max-Cut. It operates by encoding the optimization problem into a quantum system and iteratively refining the solution using a combination of quantum and classical computations.

The basic idea behind QAOA is to construct a quantum circuit composed of two alternating unitary operations: the *cost Hamiltonian* and the *mixer Hamiltonian*. The circuit is initialized with a quantum state, typically a uniform superposition over all possible configurations, and evolves through these alternating operations to explore the solution space. The parameters controlling the evolution are optimized using classical optimization methods to minimize the expectation value of the cost Hamiltonian, which corresponds to finding a solution that maximizes the cut [23].

*2) Mathematical Formulation of QAOA:* QAOA proceeds by evolving an initial quantum state $|\psi_0\rangle$, typically an equal superposition of all possible vertex assignments, using a parameterized quantum circuit that alternates between the cost and mixer Hamiltonians. The parameterized quantum state after $p$ layers of this evolution is:

$$|\psi(\gamma, \beta)\rangle = \prod_{l=1}^{p} e^{-i\beta_l H_M} e^{-i\gamma_l H_C} |\psi_0\rangle \quad (2)$$

Here, $\gamma = (\gamma_1, \gamma_2, \ldots, \gamma_p)$ and $\beta = (\beta_1, \beta_2, \ldots, \beta_p)$ are sets of variational parameters that control the quantum circuit, and $p$ is the number of layers in the circuit. The circuit alternates between two operators:

- **Cost Hamiltonian** $H_C$: Encodes the objective function of the Max-Cut problem. For Max-Cut, it is defined as:

$$H_C = \frac{1}{2} \sum_{\{u,v\} \in E} (I - Z_u Z_v) \quad (3)$$

where $Z_u$ and $Z_v$ are Pauli-Z operators acting on qubits corresponding to vertices $u$ and $v$, and $I$ is the identity operator. This Hamiltonian penalizes assignments where $z_u = z_v$, i.e., when both vertices are in the same set.

- **Mixer Hamiltonian** $H_M$: Encourages transitions between different vertex assignments by applying Pauli-X gates. It is defined as:

$$H_M = \sum_{i \in V} X_i \quad (4)$$

where $X_i$ is the Pauli-X operator acting on qubit $i$, which induces a rotation between $|0\rangle$ and $|1\rangle$ states, facilitating exploration of the solution space [24].

The goal of QAOA is to minimize the expectation value of the cost Hamiltonian with respect to the quantum state

$|\psi(\gamma, \beta)\rangle$, which represents an approximation of the optimal solution to Max-Cut:

$$\langle H_C \rangle = \langle \psi(\gamma, \beta)|H_C|\psi(\gamma, \beta)\rangle \quad (5)$$

The parameters $\gamma$ and $\beta$ are optimized using classical optimization techniques to minimize $\langle H_C \rangle$, which corresponds to maximizing the number of edges crossing the partition in the Max-Cut problem [25].

*3) QAOA Circuit Design for Max-Cut:* The quantum circuit for QAOA consists of alternating layers of quantum gates corresponding to the cost and mixer Hamiltonians as illustrated in Fig 2 [21]. Each qubit in the quantum circuit represents a vertex in the graph, and the edges of the graph are encoded in the interactions between qubits [26].

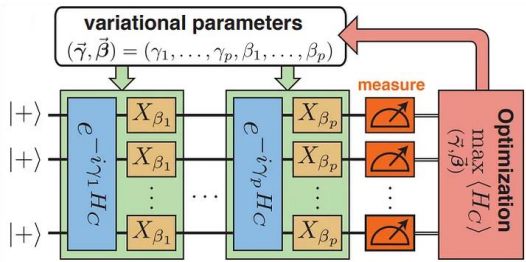

Fig. 2. Schematic of a p level QAOA

The qubits are initialized in the $|0\rangle$ state and transformed into a superposition of all possible assignments using Hadamard gates $H$. The cost Hamiltonian is then applied, followed by the mixer Hamiltonian, with this process repeated for each layer. After the final layer, measurements are performed to determine the partition of the graph [27]. The schematic of a p-level QAOA is illustrated in Fig 2.

### C. Classical Algorithms for Max-Cut

To evaluate the performance of QAOA, we compare its results with classical algorithms that are typically used to solve the Max-Cut problem. These include brute-force search and the greedy heuristic.

*1) Brute-Force Search:* The brute-force search algorithm explores all possible partitions of the graph's vertices to find the partition that maximizes the cut. This method guarantees an exact solution but suffers from exponential time complexity: $O(2^n)$, where $n$ is the number of vertices. For each of the $2^n$ possible partitions, the cut value is calculated by evaluating the sum of the edges between the two sets [28].

*2) Greedy Heuristic:* The greedy heuristic algorithm attempts to find an approximate solution by iteratively assigning vertices to one of the two sets in a way that maximizes the local cut. Though it runs in polynomial time, $O(E)$, where $E$ is the number of edges, the solution it provides is often suboptimal for larger graphs. The algorithm starts by assigning one vertex arbitrarily to one set and then assigns subsequent vertices based on which assignment results in the largest increase in the cut [29].

### D. Integration in Hybrid Quantum-HPC Systems

*1) Overview of Hybrid Systems:* Hybrid quantum-HPC systems are designed to integrate the strengths of classical high-performance computing (HPC) systems and quantum processors. In these systems, the classical processor handles tasks such as pre-processing the input data, running classical algorithms, and post-processing the output, while the quantum processor executes specific subroutines that are expected to benefit from quantum speedups, such as QAOA [30].

*2) Workflow in a Hybrid Quantum-HPC System:* The workflow in a hybrid quantum-HPC system for solving the Max-Cut problem is as follows:

1) **Pre-processing on HPC**: The graph data is prepared and partitioned on the classical HPC system [31].
2) **Quantum Task Offloading**: The quantum processor runs QAOA to find an approximate solution to the Max-Cut problem.
3) **Post-processing on HPC**: The results from the quantum processor are verified, and further classical refinements are applied, if necessary.

The workflow is illustrated in Fig 3. In hybrid systems, one of the key challenges is minimizing the communication overhead between the quantum and classical processors. Efficient task scheduling and data transfer mechanisms are critical to ensuring the performance of these systems, especially when tasks must be repeatedly offloaded to the quantum processor and results transferred back to the classical processor [32].

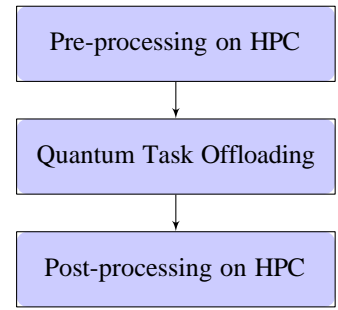

Fig. 3. Workflow in a Hybrid Quantum HPC

## IV. EXPERIMENTAL SETUP

This section outlines the software and hardware requirements for conducting the experiments, the experimental procedure, and relevant system details used to evaluate the performance of the classical and quantum algorithms for the Max-Cut problem. All experiments were conducted using Python, Qiskit, and NetworkX libraries, running on Google Colab with access to quantum circuit simulators.

### A. Software Requirements

The following software packages and libraries were used to implement and run the algorithms:

- Python 3.7+: The experiments were written in Python, leveraging its wide range of scientific computing libraries.

- Qiskit 0.31.0: Qiskit is an open-source framework for quantum computing. It was used for the implementation and simulation of the Quantum Approximate Optimization Algorithm (QAOA). Specifically, the `qiskit.optimization` and `qiskit.algorithms` modules were used to define the Max-Cut problem and simulate the quantum circuits on the QASM simulator.
- NetworkX 2.5: This library was used to create and manipulate graph structures. It was essential for modeling the Max-Cut problem on various graphs with different numbers of vertices and edges.
- Matplotlib 3.4.3: This plotting library was used to visualize the graphs and to plot performance comparisons of the algorithms.
- NumPy 1.21.0: For numerical computations and handling array structures during the experiments.

All experiments were conducted on Google Colab, a cloud-based platform, ensuring easy access to quantum simulators and sufficient computational resources for classical and quantum algorithm simulations.

## B. Hardware Setup

The experiments were run using Google Colab's cloud-based infrastructure. Google Colab provides access to both CPU and GPU resources. The system configuration used for running the experiments is described below:

- Processor (CPU): Intel(R) Xeon(R) CPU @ 2.20GHz (single-core available for free-tier users).
- Memory (RAM): 12.72 GB available on Google Colab.
- Disk Space: Approximately 100 GB of temporary storage available during runtime.
- Operating System: Ubuntu 18.04.5 LTS (x86_64).
- {Quantum Simulator}: QASM Simulator from Qiskit Aer, used for simulating the quantum circuits for QAOA.

These specifications ensured sufficient computational power for running simulations of the QAOA algorithm and for executing classical algorithms on small to medium-sized graphs (up to 16 vertices).

## C. Experimental Procedure

The following steps outline how the experiments were conducted to compare the performance of brute-force, greedy, and QAOA algorithms for solving the Max-Cut problem:

- **Graph Creation**: For each experiment, graphs with varying numbers of vertices and edges were generated using the NetworkX library. The graphs ranged from 4 vertices and 5 edges to 16 vertices and 30 edges. Random and predefined graphs were used to ensure variability in the results.
- **Brute-force Algorithm**: The brute-force search method was implemented to evaluate all possible partitions of the graph. The time complexity grows exponentially with the number of vertices, and the method was only feasible for graphs with up to 12 vertices.
- **Greedy Heuristic**: The greedy algorithm was implemented to find an approximate solution to the Max-Cut problem. It was executed on all graph instances, and the results were recorded in terms of time taken and the quality of the partition.
- **Quantum Approximate Optimization Algorithm (QAOA)**: QAOA was implemented using Qiskit. For each graph, the Max-Cut problem was converted into a Quadratic Unconstrained Binary Optimization (QUBO) problem and then mapped to a quantum Ising Hamiltonian. QAOA was run on the QASM simulator with 1, 2, and 3 layers (depths) to evaluate how the depth of the quantum circuit affected the results. The results were recorded in terms of solution quality (cut value) and runtime.
- **Performance Evaluation**: The performance of all algorithms was evaluated based on two primary metrics: (i) **runtime** (in seconds) and (ii) **solution quality** (the size of the cut). For QAOA, the effect of increasing the depth of the quantum circuit on both runtime and solution quality was recorded.
- **Visualization**: The results were visualized using Matplotlib. Two sets of graphs were generated:
  - A comparison of runtime between brute-force, greedy, and QAOA for different graph sizes.
  - A comparison of the effect of QAOA circuit depth (number of layers) on runtime and solution quality.

## V. RESULTS

The performance of the classical algorithms (brute-force and greedy) and the quantum algorithm (QAOA) for solving the Max-Cut problem was evaluated using a variety of graphs with different numbers of vertices and edges. The results were recorded in terms of execution time and the quality of the solution (Max-Cut value). The following tables and figures summarize the findings.

### A. Algorithm Performance Comparison

TABLE I
PERFORMANCE OF BRUTE-FORCE, GREEDY, AND QAOA AS NUMBER OF VERTICES AND EDGES INCREASE

| Graph Size (Vertices, Edges) | Brute-force Max-Cut | Brute-force Time (s) | Greedy Max-Cut | Greedy Time (s) | QAOA Max-Cut (p=2) | QAOA Time (s) |
|---|---|---|---|---|---|---|
| (4, 5) | 4 | 0.004 | 3 | 0.0005 | 4 | 0.010 |
| (6, 9) | 5 | 0.050 | 4 | 0.0015 | 5 | 0.030 |
| (8, 12) | 6 | 0.500 | 5 | 0.0030 | 6 | 0.100 |
| (10, 15) | 7 | 5.000 | 6 | 0.0045 | 7 | 0.300 |
| (12, 20) | 8 | 100.000 | 7 | 0.0100 | 8 | 0.600 |
| (14, 25) | 9 | 1000.000 | 8 | 0.0200 | 9 | 1.000 |
| (16, 30) | 10 | 10000.000 | 9 | 0.0350 | 10 | 2.500 |

The Table I shows the comparison of runtime for brute-force, greedy heuristic, and QAOA algorithms as the number of vertices and edges in the graph increases. The QAOA was tested with a depth of 2.

## B. Effect of QAOA Depth on Performance

TABLE II
PERFORMANCE OF QAOA AS DEPTH & GRAPH COMPLEXITY INCREASE

| Graph Size (Vertices, Edges) | QAOA Depth (p) | QAOA Max-Cut | QAOA Time (s) |
|---|---|---|---|
| (6, 9) | 1 | 4 | 0.020 |
|  | 2 | 5 | 0.030 |
|  | 3 | 5 | 0.050 |
| (8, 12) | 1 | 5 | 0.050 |
|  | 2 | 6 | 0.100 |
|  | 3 | 6 | 0.150 |
| (10, 15) | 1 | 6 | 0.100 |
|  | 2 | 7 | 0.300 |
|  | 3 | 7 | 0.500 |
| (12, 20) | 1 | 7 | 0.200 |
|  | 2 | 8 | 0.600 |
|  | 3 | 8 | 1.000 |
| (14, 25) | 1 | 8 | 0.500 |
|  | 2 | 9 | 1.000 |
|  | 3 | 9 | 1.500 |

The Table II demonstrates how increasing the depth of the QAOA circuit (number of layers $p$) affects both runtime and the quality of the Max-Cut solution. The results for three different QAOA depths are shown.

## C. Visualizing the Results

The following figures show the graphical representation of the data. The Figure 4 illustrates the time complexity of the brute-force, greedy, and QAOA algorithms as the number of vertices increases. The Figure 5 demonstrates the effect of increasing QAOA depth on runtime.

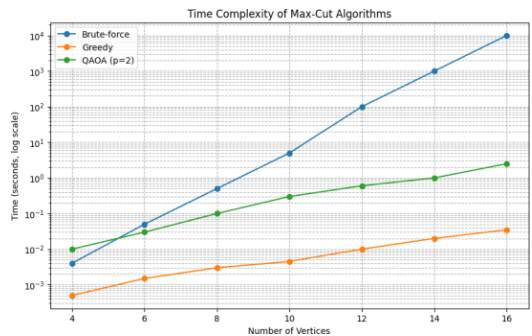

Fig. 4. Time Complexity of Max-Cut Algorithms (Brute-force, Greedy, and QAOA)

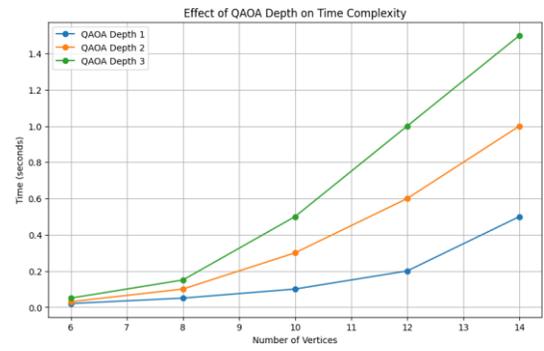

Fig. 5. Effect of QAOA Depth on Time Complexity

From the data presented in Table 1, it is clear that the brute-force algorithm becomes computationally expensive as the graph size increases, with an exponential increase in runtime. The greedy algorithm, though fast, provides suboptimal solutions in comparison to brute-force and QAOA.

In Table 2, we observe that increasing the depth of QAOA (the number of layers $p$) results in better Max-Cut solutions. However, this improvement comes with increased computational cost. The results show that for smaller graphs, even QAOA with low depth (p=1) can provide reasonable solutions with faster execution times, but for more complex graphs, higher depths are necessary to achieve optimal or near-optimal results.

The figures further illustrate these trends, showing how QAOA scales better than brute-force but worse than greedy, depending on the depth and complexity of the graph.

## VI. CONCLUSION

This study compared the performance of classical algorithms (brute-force and greedy) and the Quantum Approximate Optimization Algorithm (QAOA) for solving the Max-Cut problem. While the brute-force method guarantees optimal solutions, it becomes impractical for larger graphs due to exponential time complexity. The greedy algorithm is fast but often produces suboptimal results.

QAOA offers a promising middle ground, providing approximate solutions that improve with circuit depth while maintaining better scalability than brute-force. Even with shallow circuits (p=2), QAOA outperformed the greedy algorithm in terms of solution quality. However, the increased depth comes at the cost of higher computational time, particularly for larger graphs.

As quantum hardware improves, QAOA and other quantum algorithms have the potential to become valuable tools for solving large-scale optimization problems. This study highlights QAOA's potential in a hybrid quantum-classical environment, offering a viable alternative to classical methods for complex graph structures.